\documentclass[twocolumn]{aastex63} 

\usepackage[utf8]{inputenc}
\usepackage[caption=false]{subfig}
\usepackage{natbib}
\usepackage{comment} 
\usepackage{csvsimple, array, longtable, booktabs, amsmath, amssymb, bm, hyphenat, etoolbox} 

\usepackage{hyperref}
\usepackage[all]{hypcap} 
\hypersetup{ 
                colorlinks=true, %
                breaklinks=true, %
                pdfauthor={Maurice L. Wilson}, %
                pdftitle={Model Solar to Stellar CME }, %
                citecolor=blue, %
                linkcolor=blue %
                }

\newcommand{\rsun}{\ensuremath{ R_\odot }}
\newcommand{\Ang}{\ensuremath{ \mathrm{\AA} }}
\newcommand{\unitn}{cm\ensuremath{ ^{-3} }}
\newcommand{\unitv}{km~s\ensuremath{ ^{-1} }}
\newcommand{\ionn}[2]{#1~{\footnotesize #2}}
\newbool{egshow}\setbool{egshow}{true}\ifbool{egshow}{ \newcommand{\egcite}{e.g.,} }{ \newcommand{\egcite}{} }
\newcommand{\scinot}[2]{#1\ensuremath{\times}10\ensuremath{^{#2}}}

\newcommand{\affilHarvardCfA}{Center for Astrophysics, Harvard \& Smithsonian, 60 Garden St, Cambridge, MA 02138, USA}
                
\begin{document}

\title{Solar CME Plasma Diagnostics Expressed as Potential Stellar CME Signatures}
\shorttitle{Potential Stellar CME Spectroscopic Signatures}

\shortauthors{Wilson and Raymond}


\author[0000-0003-1928-0578]{Maurice L. Wilson}
\affiliation{\affilHarvardCfA}

\author[0000-0002-7868-1622]{John C. Raymond} 
\affiliation{\affilHarvardCfA}

\correspondingauthor{Maurice L. Wilson}
\email{maurice.wilson@cfa.harvard.edu}


\begin{abstract}

Solar coronal mass ejections (CMEs) have a strong association with solar flares that is not fully understood.  This characteristic of our Sun's magnetic activity may also occur on other stars, but the lack of successfully detected stellar CMEs makes it difficult to perform statistical studies that might show a similar association between CMEs and flares.  Because of the potentially strong association, the search for stellar CMEs often starts with a successful search for superflares on magnetically active stars.  Regardless of the flare's presence, we emphasize the utility of searching for CME-specific spectroscopic signatures when attempting to find and confirm stellar CME candidates.  We use solar CMEs as examples of why a multitude of ultraviolet emission lines, when detected simultaneously, can substantially improve the credibility of spectroscopically discovered stellar CME candidates.  We make predictions on how bright CME-related emission lines can be if they derived from distant stars.  We recommend the use of three emission lines in particular (\ionn{C}{IV} 1550~\Ang, \ionn{O}{VI} 1032~\Ang, and \ionn{C}{III} 977~\Ang) due to their potentially bright signal and convenient diagnostic capabilities that can be used to confirm if an observational signature truly derives from a stellar CME. 

 

\end{abstract}

\section{Introduction}

On the solar surface, magnetized plasma frequently experiences vehement instabilities within local magnetic field structures that spawn the eruption of coronal mass ejections (CMEs).  Since their original discovery, a plethora of CMEs have been studied, but the mechanisms that drive the initial eruption are still not fully understood \citep{Hansen.1971, Tousey.1973, Gosling.1974}.  
However, much progress has been made on this front as a variety of observational techniques have been used to give unique perspectives on these transient events.  Imagers monitoring the solar disk in high-energy bandpasses have worked well with white light coronagraphs when attempting to evaluate the CME kinetic energy.  Such observations have shown that CME velocities can range from a few tens to a few thousand~\unitv\ while its accumulating mass is typically found to be in the range $10^{14-16}$~g.  As the CME traverses the corona, its physical conditions can be deduced from spectroscopic measurements.  Its heating and cooling processes dictate the temperature for the relatively cool material at 10$^4$~K and hot material at 10$^7$~K.  

Such physical properties inform the thermal energy, kinetic energy, and other components of the CME energy budget but can vary for distinct structures within the CME.  The three commonly recognized CME parts that are adjacent but exhibit distinct properties are the leading edge, the flux rope, and the prominence core \citep{Illing.1985}.  The energy budget is distributed throughout these features and originates primarily from the magnetic energy released upon eruption.  This magnetic energy is liberated through a series of magnetic reconnections that reconfigure the magnetic structures that typically reside above flare loops. 

A complex arcade of many magnetic loops can form two parallel structures with their chromospheric footpoints and yield the two-ribbon flare that resides below a current sheet.  This current sheet connects the (pre-eruption) flux rope and prominence to the surface flare loops.  Such a flare can brighten in H$\alpha$ for several minutes, reach its peak, and take hours for its radiative energy to decay as its heating remains somewhat steady \citep{Carmichael.1964, Sturrock.1966, Hirayama.1974, Kopp.1976, Tsuneta.1992.flare}.  Conversely, a compact flare comprises only one or a few magnetic loops that become unstable and are impulsively heated.  Compared to the two-ribbon flare, the compact flare's profile has a faster rise and decay while also emitting less radiation \citep{Alfven.1967, Pallavicini.1977, Dennis.1989, Masuda.1994}.

Solar flares do not always occur with solar CMEs, but there is evidence to suggest that both phenomena become more coupled as the flare energy increases, which then corresponds to an increasing CME kinetic energy \citep{Yashiro.2009, Aarnio.2011}.  Solar flares radiate energies of $10^{28-32}$~ergs and CMEs exhibit kinetic energies of $10^{28-33}$~ergs \citep[\egcite][]{Emslie.2012, Aschwanden.2017}.  Their coinciding occurrence at high energies, particularly when the two-ribbon flare is involved, may be the consequence of their common place of origin when both phenomena commence under the same magnetic fields of high pressure and complexity.  

Between both forms of solar magnetic activity, it is not clear if their empirically derived coupling at high energies is valid for other stars.  For stellar magnetic activity, statistical surveys on flare-CME relations are not feasible due to the dearth of stellar CME candidates detected.

Evidence for the detection of a stellar CME candidate is bolstered when accompanied by a stellar flare.  Since their initial discoveries \citep{Hertzsprung.1924}, stellar flares have been observed to have characteristics similar to solar flares \citep{Hawley.1995, Guedel.1996}.  Thus, there is a chance that the processes initiating stellar CMEs are similar to solar CMEs.  Furthermore, correlations with flare-CME occurrence rates as seen from the Sun may also be present on other stars.  However, these solar to stellar extrapolations cannot be verified until many stellar CME candidates are found, confirmed, and applied to statistical surveys.  Thus far, the candidates are plagued with large uncertainties due to the lack of spatial, temporal, or spectral resolution and therefore require significant assumptions to interpret the features as CMEs.

Favorable geometric assumptions have been invoked to infer the presence of stellar CME candidates from ultraviolet dimmings. \cite{Giampapa.1982} acquired high-cadence photometry in the Johnson \textit{U}~bandpass during a flare event from the binary system EQ~Pegasi, which is at a distance of 6.2~pc away and consists of two M~dwarf stars that are separated by at least 24~AU.  Their light curve showed a 25\% dip in the quiescent flux that lasted almost 5~minutes.  This preceded a flare that lasted almost 20~minutes.  They suggested that this ``pre-flare dip'' might be due to mechanisms that are similar to phenomena seen in solar filament eruptions.  As a filament destabilizes, some of the frozen-in plasma of magnetic loops can flow downward under the force of gravity.  This plasma transfers its kinetic energy to the underlying chromospheric material and contributes to the onset of a flare brightening via collisional excitation \citep[\egcite][]{Hyder.1967.July, Hyder.1967.Nov}.  During eruption, other rising filament material can travel at speeds near 100~\unitv.  Consequently, this plasma can vanish in solar disk observations where its emission is Doppler-shifted out of an observation's narrowband filter and thus commences a pre-flare dimming event for a given bandpass.  In the case of the EQ~Pegasi flare, \cite{Giampapa.1982} suggested that their pre-flare dip may have been the result of a prominence eruption on the limb where some of the prominence material travels at an angle that causes it to overlap with the face of the (unresolved) stellar disk.  Therefore, the 25\% dip was perhaps due to the relatively cool prominence (or filament) material temporarily obscuring a quarter of the stellar disk.  

While investigating another stellar flare, \cite{Ambruster.1986} also measured a dip in their light curve that could be due to the geometry of stellar CME event.  However, they suggested that the typical conclusions drawn for pre-flare dips are likely not applicable because their observations indicated a dip that occurred after the flare with the dip lasting much longer than 5~minutes.  They used spectra from the International Ultraviolet Explorer (IUE) to investigate a flare from the M4.5V star, EV Lacertae, at a distance of 5.1~pc away.  For exposure times of 45~minutes, the spectra for ultraviolet emission lines, such as the \ionn{C}{IV} 1550~\Ang\ line and the \ionn{Mg}{II} 2800~\Ang\ line, formed light curves that revealed a dimming that occurred almost an hour after a flare event and lasted for almost 1.5~hours.  \cite{Ambruster.1986} favored a CME conclusion to explain the long, delayed dimming.  
During the CME's travel and expansion, its prominence core would have absorbed the \ionn{C}{IV} and \ionn{Mg}{II} radiation coming from the stellar surface.  
An analogous solar phenomenon occurs when solar filament eruptions are observed as absorption features in ultraviolet images of the solar disk \citep[\egcite][]{Filippov.2002, Kundu.2004}.  However, the obscuring of the hot coronal material below the cool eruptive filament typically yields a negligible dimming when considering the total integrated flux of the solar disk, i.e. an unresolved disk. 

Doppler blueshifts and spectral line asymmetries have also been used to propose stellar CME candidates.  Such techniques were used by \cite{Houdebine.1990}, \cite{Argiroffi.2019}, and \cite{Namekata.2021} to introduce their plausible candidates, but many other stellar CME candidates also give unique incentives for scrutinizing each CME candidate's magnetically active host star as the resolution and precision of modern instruments continues to improve.

\cite{Houdebine.1990} used the European Southern Observatory (ESO) to investigate an impulsive flaring event on the M4.5V star, AD~Leonis (AD~Leo), 5.0~pc away.  During the flare, enhancements in the blue wings of the H$\gamma$ and H$\delta$ Balmer line profiles were seen.  Over 1-minute exposure times, the line asymmetry indicated a Doppler velocity of 5830~\unitv\ and later 3750~\unitv.  In a separate analysis, \cite{Leitzinger.2011} studied two flares from AD~Leo with spectra from the Far Ultraviolet Spectroscopic Explorer (FUSE) \citep{Christian.2006}.  For exposure times ranging from a few minutes to about 30~minutes, the prominent features derived from \ionn{C}{III} emission lines at 977 and 1176~\Ang\ as well as \ionn{O}{VI} emission lines at 1032~\Ang\ and 1038~\Ang.  A blueshift in the \ionn{O}{VI} 1032 emission line implied a Doppler velocity of 84~\unitv\ while the other lines conveyed substantially slower Doppler velocities.    

\cite{Argiroffi.2019} used Chandra to study the flaring of the G1III star, OU~Andromedae (OU~And), 139.5~pc away.  The duration of the flare's rise and decay lasted for 40~ks, and the spectra conveyed significant redshifts and blueshifts.  Blueshifts of several hundred~\unitv\ were seen during the flare's rising phase and were interpreted as the motion of heated chromospheric plasma within a flare loop.  Evidence for a CME was seen in a blueshift of 90~\unitv\ that occurred after the flare, over an integration time of 58~ks.  

\cite{Namekata.2021} introduced a stellar CME candidate in their study of a flaring event on the G1.5V star, EK Draconis (EK Dra), 34.4~pc away.  The flare was detected with the optical photometry of the Transiting Exoplanet Survey Satellite (TESS) while the spectroscopic instruments on the Seimei and Nayuta telescopes monitored the flare's H$\alpha$ profile \cite{Ricker.2014, Kurita.2020}.  The impulsive flare lasted for 16~minutes and coincided with the brightening of redshifted H$\alpha$ emission.  Post-flare, the H$\alpha$ line exhibited blueshifted absorption signatures that lasted for at least 1.5~hours.  During this time, the spectra were acquired with exposure times of either 30~seconds or 3~minutes, and an initial blueshift of 510~\unitv\ was followed by a series of decelerating blueshifts.  This was interpreted as the presence of relatively cool \ionn{H}{I} plasma from a stellar filament eruption traveling toward the observer for almost two hours.

These exemplary candidates (and many more) are interpreted as potential stellar CMEs primarily because their detected signals are analogous in some way to the observational signatures of solar CMEs \citep[\egcite][]{Osten.2017, Moschou.2019, Vida.2019}.  However, without spatially resolving the star, these candidates require that assumptions be made about the morphology of the CME.  Furthermore, the plasma diagnostics are often not determined by a variety of spectral lines, which would help constrain the travelling plasma's ever-changing physical conditions.  Among many parameters, the candidates have estimates of mass, absolute velocity, temperature, density, and ionization states that are largely uncertain.  Consequently, each candidate's identity as a stellar CME remains uncertain.    

If more spectral lines are utilized simultaneously in the hunt for stellar CMEs, deducing the plasma properties would require less assumptions.  Considering this, we emphasize the benefits of using observationally constrained plasma diagnostics from solar CMEs and we make predictions for how bright the same CME signal would be for certain spectral lines if observed as stellar CMEs.  In this work, we scrutinize three previously studied solar CMEs and determine the feasibility of detecting the same spectral signal for the generic case of a stellar CME producing emission lines in the ultraviolet wavebands.  

For our predictions, we assume the hypothetical emitting plasma is more massive than the most massive solar CMEs because many stellar flares have been observed to be more energetic than the most luminous solar flares.  Thus, we assign a mass of 10$^{17}$~g for the stellar CME.  
For an emission line observed during a solar CME, we calculate the luminosity and amplify it by a factor corresponding to the ratio between the solar CME's mass and our assumed stellar CME mass, which can be up to a factor of 10$^3$.  Similarly, this factor can also correspond to how the observed energy of stellar superflares ($>10^{33}$~ergs) can be thousands of times greater than many solar flares.  We assume that the scaled up luminosity yields a flux that will be detected by an instrument that is subject to an effective area similar to that of the \textit{Hubble Space Telescope's} (\textit{HST}) Cosmic Origins Spectrograph (COS).

For three distinct solar CMEs, the calculations are described in Sections \S\ref{sect: event Landi}, \S\ref{sect: event Ciaravella}, and \S\ref{sect: event Wilson}.  We give our final remarks in \S\ref{sect: conclusion} regarding important caveats that must be considered when searching for stellar CMEs, including the utility (or futility) of assuming solar-like properties to make sense of constrained (or unconstrained) properties of stellar magnetic activity signatures.

\vspace{20pt}

\section{Solar CME Event, 9 April 2008}\label{sect: event Landi}

We consider the CME studied by \citet{Landi.2010}, which erupted on 9 April 2008.  They examined the data acquired by instruments on the \textit{SOHO} \citep{Domingo.1995}, \textit{Hinode} \citep{Kosugi.2007}, and \textit{STEREO} \citep{Kaiser.2005} spacecrafts.  To gather spectra, the slit apertures of \textit{Hinode}/EIS \citep{Culhane.2007} and \textit{SOHO}/UVCS \citep{Kohl.1995} were monitoring the corona at heliocentric distances of 1.1~\rsun\ and 1.9~\rsun\ respectively.  During the initial eruption, photometry of the solar disk was captured by \textit{Hinode}/XRT \citep{Golub.2007}, \textit{SOHO}/EIT \citep{Delaboudiniere.1995}, and \textit{STEREO-A}/EUVI; and, this was complemented by the coronagraph imagers \textit{SOHO}/LASCO \citep{Brueckner.1995} and \textit{STEREO-A}/SECCHI/COR~1 and COR~2.  As a result, the cumulative spatial coverage for which this event was studied ranged from the solar disk out to 22~\rsun.

\subsection{Solar CME Characteristics}

The CME was initially seen near the southwest limb of the solar disk.  The CME's leading edge was only visible off the limb within the \textit{STEREO-A} photometry.  The flux rope was too faint and thus observed by none of the instruments, while the CME core was sufficiently bright for all of the instruments to capture it.  Consequently, the physical properties deduced were based solely on the observed core and leading edge.

The total mass was roughly $10^{14}$~g.  The leading edge was consistently accelerating as its velocity approached 700~\unitv\ near 3~\rsun, which is where the plasma became too dim to track any further.  The core material's acceleration persisted until the plasma reached 5~\rsun, which is where its velocity reached 475~\unitv\ and remained constant out to 22~\rsun.  The velocity estimates were derived from the observed trajectory of the CME as it travelled across the imagers' plane of sky (POS) and created Doppler shifts in the spectrometers' spectra via the CME's motion along the line of sight (LOS). 

The physical conditions experienced by the CME were determined primarily from the \textit{Hinode}/EIS spectra at 1.1~\rsun. Only the CME core was detected, and its density was found from density-sensitive intensity ratios between the emission lines observed by EIS.  Assuming ionization equilibrium, the temperature was estimated for the ions detected and used to model the thermal distribution of the CME core plasma.  The emission line ratios indicated the presence of various plasma environments within the core volume that experience distinct densities and temperatures.  The densities evaluated were in the range log~$n_e$~[\unitn] = 7.75--11.3, and the thermal distribution conspicuously revealed two distinct temperature ranges for the core material: log~$T_e$~[K] 4.9--5.4 and 5.5--5.9.  Presumably the CME expanded and cooled as it travelled beyond 1.1~\rsun, but there were no detectable density- nor temperature-sensitive line ratios to verify the evolving plasma diagnostics.

\subsection{Solar to Stellar Extrapolation}\label{sect: event Landi extrapolation}

The large range of spectral lines and ions detected by EIS allowed for a large range in temperature to be found in the CME core's thermal distribution.  This distribution was given in the form of the differential emission measure (DEM), which describes how the amount of plasma emitting the observed radiation changes with the temperature.  We use the DEM of this CME event to estimate the luminosity given off for a few ultraviolet spectral lines, and we consider if certain emission lines would be detectable if the CME's brightness derived from a distant star.   

For each emission line that we consider, we estimate the plasma's emission measure (EM) and convert it to an absolute intensity by applying the emissivity model of \cite{Raymond.1981}.  With the temperature-dependent DEM curve given by \cite{Landi.2010}, we integrate the DEM over a small interval of temperatures (0.1 dex) near the emitting ion's temperature of maximum formation ($T_{max}$).  This yields the EM for each spectral line of interest.  

We chose spectral lines that were already used to constrain the model of \cite{Raymond.1981}, which introduces a proportionality between the emission measure and absolute intensity:  
\begin{equation}
    I_{EM} = \frac{\rm{EM}}{10^{26}\ [\rm{cm}^{-5}] }  I_{RD},
\end{equation}
where the predicted intensity $I_{EM}$ for a given emission line is proportional the line's empirically derived intensity $I_{RD}$ from \cite{Raymond.1981}, which corresponds to an emission measure of $10^{26}$~cm$^{-5}$.

Once the model intensity $I_{EM}$ is calculated for an observationally constrained EM, we convert the intensity into the following photon luminosity for any given transition line,
\begin{equation}
    L_{\odot, CME}\ [\textrm{photons\ s}^{-1}]  = 4\pi r^2 I_{EM},
\end{equation}
where we find the emitting plasma's radius to be approximately $R$=0.105~\rsun\ over the POS.  The area roughly encompasses the size and shape of the CME core as seen from the \textit{Hinode}/XRT images presented by \cite{Landi.2010}.

For this solar CME luminosity, we estimate the flux that might be observed if the same CME was successfully launched from a distant star. The detected flux is expressed in the simplified form,
\begin{equation} \label{eq: flux}
    F_{\star, CME} \ [\textrm{photons}] \sim \frac{ L_{\odot, CME} \cdot f_\star}{ 4\pi d^2_\star } \cdot A_{eff} \cdot \Delta t,
\end{equation}
which considers a star of distance $d_\star$ away, a telescope with an effective area of $A_{eff}$, an integration time of $\Delta t$, and a factor of $f_\star$ to account for a plausible discrepancy between typical solar CMEs and stellar CMEs of magnetically active stars.  The flux corresponds to the generic case of an ultraviolet instrument with a wavelength-dependent $A_{eff}$ that observes a stellar CME from $d$=10~pc away that is a factor $f_\star$ times brighter than the reference solar CME.  For the CME examined by \cite{Landi.2010}, we scale the luminosity up by a factor of $f_\star= 10^3$, which corresponds to the ratio of our assumed stellar CME mass and this solar CME mass.  If $\Delta t$=3600~seconds and the data gathered over an hour of observations are co-added, the resultant fluences are as shown in Table~\ref{table: Landi CME}. 

Among the emission lines tested, the \ionn{C}{IV} line gives the greatest signal and results in a signal-to-noise ratio of S/N~$\approx$~85 if only poisson photon noise is assumed.  The evaluation of this signal employs generic assumptions that do not account for all of the specific sources of astrophysical and instrumental noise that could hinder the detection of this predicted signal.  A detailed study of a predicted stellar CME's signal for a specific detector's systematic noise is beyond the scope of this paper; but, such an in-depth study would benefit from specifically considering the M~dwarf flare star Proxima Centauri as a promising candidate for launching detectable stellar CMEs.  Being the closest star system to our solar system, its distance of $d$=1.30~pc away results in a factor of $\sim$60 increase to our flux approximations.  Consequently, the stellar CME signal from the \ionn{C}{IV} emission line would give a S/N~$\approx$~640.  In this case, a shorter integration time may suffice.

\newcommand{\tsize}{\footnotesize}

\capstartfalse
\begin{table*}

    \begin{longtable}{ccccccccccc} 
        
        \label{table: Landi CME}
        \\\multicolumn{11}{c}{ {\scshape \tablename\ \thetable}: Solar ultraviolet emission lines considered for stellar CME detection }
        \\ \midrule\midrule
  
        \tsize $\lambda$  &\tsize  Ion  &  \multicolumn{3}{c}{Transition\tsize}  &\tsize  log $T_{max}$  &\tsize  log DEM  &\tsize $I_{RD}$  &  $A_{eff}$  &\tsize  $F_{10\rm{pc},CME}$  &\tsize  $F_{Proxima, CME}$  \\ 
        \ 
        \tsize [$\mathrm{\AA}$]  &   &   &   &   &\tsize [K] &\tsize  [cm$^{-5}$~K$^{-1}$]$^b$  &\tsize  [erg~s$^{-1}$~cm$^{-2}$~sr$^{-1}$]$^c$ & [cm$^2$]$^d$ &\tsize [photons]$^e$  &\tsize  [photons]$^e$  \\\midrule
        
        \tsize 1663.5  &\tsize  \ionn{O}{III}  &\tsize  $2\rm{p}^2\ \ ^3\rm{P}_{1,2}$  &-&\tsize $2\rm{s}2\rm{p}^3\ \ ^5\rm{S}_{2}$   &\tsize  5.0  &\tsize  20.8  &\tsize     87.9 & 400&\tsize  86  &\tsize  \scinot{5.1}{3} \\
        \tsize 1550$^a$  &\tsize  \ionn{C}{IV}  &\tsize  $2\rm{s}\ \ ^2\rm{S}_{1/2}$ &-&\tsize $2\rm{p}\ \ ^2\rm{P}_{1/2,3/2}$ &\tsize  5.0  &\tsize  20.8  &\tsize \scinot{3.40}{3}  & 900  &\tsize  \scinot{7.0}{3}  &\tsize  \scinot{4.1}{5}  \\ 
        \tsize 1484.9 &\tsize  \ionn{N}{IV}  &\tsize $2\rm{s}^2\ \ ^2\rm{S}_0$ &-&\tsize $2\rm{s}2\rm{p}\ \ ^3\rm{P}_{1,2}$ &\tsize  5.1  &\tsize  21.0  &\tsize 69.2  & 1200   &\tsize  365  &\tsize  \scinot{2.2}{4} \\
        \tsize 1718.6  &\tsize  \ionn{N}{IV}  &\tsize $2\rm{s}2\rm{p}\ \ ^1\rm{P}_1$ &-&\tsize $2\rm{p}^2\ \ ^1\rm{D}_2$ &\tsize  5.1  &\tsize  21.0  &\tsize 8.53  & 300 &\tsize 13  &\tsize  768 \\
        \tsize 1238.8  &\tsize  \ionn{N}{V}  &\tsize  $2\rm{s}\ \ ^2\rm{S}_{1/2}$  &-&\tsize $2\rm{p}\ \ ^2\rm{P}_{3/2}$ &\tsize  5.3  &\tsize  20.2  &\tsize \scinot{4.75}{2}  & 2400   &\tsize  \scinot{1.0}{3}  &\tsize  \scinot{6.2}{4} \\
        \tsize 1031.9  &\tsize  \ionn{O}{VI}  &\tsize   $2\rm{s}\ \ ^2\rm{S}_{1/2}$ &-&\tsize  $2\rm{p}\ \ ^2\rm{P}_{3/2}$  &\tsize  5.5  &\tsize  19.0  &\tsize \scinot{1.44}{3}   & 20   &\tsize  2  &\tsize  131 \\
        
        \midrule
        \multicolumn{11}{l}{\tsize$^a$At 1550~\Ang, a blended emission line is observed if two spectral lines of \ionn{C}{IV}, at 1548.2 and 1550.8~\Ang, are not spectroscopically resolved. }
        \\\multicolumn{11}{l}{\tsize$^b$DEM is for the CME studied by \cite{Landi.2010}. }
        \\\multicolumn{11}{l}{\tsize$^c$The $I_{RD}$ is from \cite{Raymond.1977}. }
        \\\multicolumn{11}{l}{\tsize$^d$The $A_{eff}$ from the \textit{HST}/COS gratings is merely a realistic reference we apply to our predicted signal \citep{Fischer.2021}. }
        \\\multicolumn{11}{l}{\tsize $^e$The detected photons (cf. Equation~\ref{eq: flux}) account for the assumed $A_{eff}$.
        }

    \end{longtable}
\end{table*}
\capstarttrue

\section{Solar CME Event, 12 December 1997}\label{sect: event Ciaravella}

Another CME event that we consider occurred during 12 December 1997 \citep{Ciaravella.2000, Ciaravella.2001}.  Images of the CME's footpoints and eruption from the \textit{Yohkoh} Soft X-ray Telescope (SXT) and the Meudon Observatory H$\alpha$ photometry complemented the data acquired by three \textit{SOHO} instruments: EIT, LASCO, and UVCS.  \cite{Ciaravella.2000} used the pre-CME solar disk images from SXT and Meudon to monitor the evolution of the active regions near the CME's launch site \citep{Tsuneta.1991, Duff.1982}.  The CME's morphology and physical conditions were determined from the EIT, LASCO, and UVCS post-eruption observations.

\vspace{20pt}

\subsection{Solar CME Characteristics}

Upon eruption, only the prominence core component of the CME was clearly discerned in the \textit{SOHO} fields of view.  In EIT's 195~\Ang\ filter, it could be seen up to 1.2~\rsun.  \cite{Ciaravella.2000} suggested that it was mostly cool ions, such as \ionn{O}{IV} at \scinot{1.5}{5}~K or \ionn{O}{V} at \scinot{2.5}{5}~K, that were captured in the images.  This is in spite of there usually being hot \ionn{Fe}{XII} at \scinot{1.3}{6}~K that dominates the 195~\Ang\ bandpass of EIT.  The prominence material was seen off the northwest limb traveling at 140~\unitv\ near 1.7~\rsun\ and later seen traveling at more than 200~\unitv\ in the LASCO/C3 field of view.  The Doppler shifts from the UVCS spectra conspicuously indicate the plasma's helical motion, which is similar to the behavior theorized by CME flux rope models \citep[\egcite][]{Gibson.1998, Guo.1998}.  

The physical conditions were deduced from the UVCS spectra acquired when the slit was positioned at 1.7~\rsun.  A temperature-sensitive intensity ratio from two Si lines (at 1303 and 1206~\Ang) was used to confirm the plasma's state of ionization equilibrium.  Thus, the temperatures of maximum ion formation were utilized for the plasma diagnostics and ranged from $4.2<$~log~$T_e$~[K]~$<5.5$.  With these temperatures, the ions' emission measures were evaluated from the spectral lines detected and the fiducial atomic transition rates \citep{Raymond.1977, Scholz.1991, Griffin.1993}.  
For almost all of the ions observed, \cite{Ciaravella.2000} found a flat emission measure distribution across their range of temperatures.  The column emission measure can be approximated as EM~$\approx \int n^2_e dl$, which integrates over the observed plasma's column depth along the LOS.  Strands of filamentary prominence material were seen crossing the UVCS slit, and the width of their presumably cylindrical structures were estimated as the column depths.  \cite{Ciaravella.2001} inferred the densities from the EMs and column depths which were in the range $6.0<$~log~$n_e$~[\unitn]~$<7.5$. The volume was estimated from the aforementioned LOS column depth and the prominence material's POS area as seen from LASCO.  Consequently, the mass of the structures ranged from roughly 10$^{13}$ to 10$^{14}$~g.

\subsection{Solar to Stellar Extrapolation}\label{sect: event Ciaravella extrapolation}

We primarily consider the EIT observations of the CME prominence material when extrapolating the flux out to stellar distances.  We use the detected count rates from the CME and apply the EIT instrumental response function as summarized by \cite{Delaboudiniere.1995} in Figure~9 of their work. As a result, we estimate the CME's volume emission measure (EM$_V \approx \int n^2_e dV$) with the following relationship,
\begin{equation}\label{eq: eit response}
    \frac{ \textrm{EM}_{V, \rm{CME}} }{ 10^{44}\ \rm{cm}^{-3} } = \frac{ \mathcal{L}_{\odot, \rm{CME}} }{ \mathcal{L}_{\rm{EIT}} },
\end{equation}
which utilizes EIT's signal $\mathcal{L}_{\rm{EIT}}$ in CCD counts per second as a function of temperature when the volume emission measure is $10^{44}$~cm$^{-3}$.  EIT observations can detect plasma of temperatures within the range 0.06--3~MK.  At a temperature of 10$^5$~K, the instrumental response is reported to be $\mathcal{L}_{\rm{EIT}} [\rm{counts~s}^{-1}] = 2.0$.  At the same 10$^5$~K temperature, the plasma will likely be bright in the \ionn{C}{IV} emission line that is discussed in \S\ref{sect: event Landi extrapolation}.  We estimate the average $\mathcal{L}_{\odot, \rm{CME}}=$ \scinot{4.2}{3}~counts~s$^{-1}$ from the EIT images of the prominence material and, with Equation~\ref{eq: eit response}, find that EM$_{V, \rm{CME}} = \scinot{2.1}{47}$~\unitn.  This excludes the image (at the time of 23:34 UTC) that is the first to expose the erupting prominence material but is likely spatially coherent with bright flare material. 

Relatively cool prominence material typically has a temperature of about $10^5$~K upon eruption and is seen at a similar temperature later in the the mid-to-high corona.  This implies that there must be a substantial amount of heating within the plasma that is capable of being balanced with the CME's radiative and expansion cooling \citep{Akmal.2001, Lee.2009, Murphy.2011, Rivera.2019.April}.  Thus, the radiative loss rate of the emitting plasma can act as a lower limit of the heating rate that levels out (or increases) the plasma temperature and therefore yields a corresponding emission measure.  At the temperature of maximum ion formation for \ionn{C}{IV}, we use its radiative loss rate coefficient, $\Lambda_{\rm{C\ IV}}$(T=10$^{5.0}$~K)$ = \scinot{5.4}{-8} \ \frac{ \rm{photons\ cm}^3 }{ s }$ from the CHIANTI database \citep{Dere.1997, Dere.2019}.  We characterize the luminosity as 
\begin{equation}
    L_{\odot, CME}\ [\textrm{photons\ s}^{-1}] = A_Z f_{Z,z} \Lambda (T) \times \textrm{EM}_{V, \rm{CME}}
\end{equation}
which includes the (coronal) elemental abundance $A_Z$ and ionization fraction $f_{Z,z}$ for ion $z$ of element $Z$ \citep{Feldman.1992}.  We find that $L_{\odot, CME} =$ \scinot{1.2}{36}~photons~s$^{-1}$ for \ionn{C}{IV}. 

With this luminosity, we apply Equation~\ref{eq: flux} with $f_\star = 10^3$ and estimate the expected stellar fluences.  We find $F_{10\rm{pc}, CME}$ and $F_{Proxima, CME}$ equate to \scinot{5.2}{3} and \scinot{3.1}{5}~counts which suggests the S/N~$\approx$ 70 and 555, respectively, if the integration time is only 1~minute.  Since these estimates are heavily dependent on the EIT's instrumental response function, the detected signal and the required exposure time may vary drastically for the capabilities of a different instrument.

As a comparison, the \ionn{O}{VI} 1032~\Ang\ line is also worth considering since it is typically one of the brightest spectral lines observed when UVCS spectra are taken of the ambient corona and the transient coronal mass ejections.  The instrumental response is at most 5.0~counts~s$^{-1}$ for plasma at $T_e=10^{5.5}$~K. The radiative loss rate coefficient is $\Lambda_{\rm{O\ VI}}(T_e=10^{5.5}$~K)$ = \scinot{3.1}{-8} \ \frac{ \rm{photons\ cm}^3 }{ s }$, which results in the following predicted fluxes for detection over an integration time of 1~minute: $F_{10\rm{pc}, CME}$ and $F_{Proxima, CME}$ are 48 and \scinot{2.9}{3}~counts which imply a S/N~$\approx$ 5 and 50, respectively.  In this case, a longer exposure time may be preferable when attempting to detect a stellar CME signal from the \ionn{O}{VI} 1032~\Ang\ line.

\section{Solar CME Event, 17 May 1999}\label{sect: event Wilson}

The first CME of interest (cf. \S\ref{sect: event Landi}) was analyzed primarily from its spectra at 1.1~\rsun.  The same is true for the second CME of interest (cf. \S\ref{sect: event Ciaravella}) at 1.7~\rsun.  We now pay heed to a CME event where its detailed spectral information was gathered at 2.6 and 3.1~\rsun.  This CME event was observed on 17 May 1999 by \textit{SOHO}'s EIT, LASCO, and UVCS \citep{Wilson.2022}.

\subsection{Solar CME Characteristics}

Upon eruption off the northwest limb, strands of the prominence material were captured by EIT.  In the white light images of LASCO/C2, the canonical three-part CME can be discerned.  Evidently, beyond 3~\rsun, the leading edge accelerated to 500~\unitv\ and was followed by a featureless void that separated it from the large, amorphous prominence core.  The column density of the CME features discerned along the LASCO's plane of sky suggests a mass of 10$^{15}$~g.

The automated observing program of UVCS serendipitously placed the slit aperture at heights in the corona that were along the CME's unpredictable path.  Only the prominence core was seen crossing the slit's field of view at any given exposure.  Coincidentally, some specific plasma structures within the core were seen crossing the slit twice---once at 2.6~\rsun\ and once at 3.1~\rsun.  Thus, the average POS velocity was determined between the two heights along with the LOS velocity from the Doppler shifts of the spectra.  Together, the absolute velocity was found to be 250~\unitv\ for the prominence core. The plasma diagnostics were only evaluated for the core material featured in the UVCS spectra.  

Due to the many spectral lines detected, a variety of observed intensity ratios were modelled to estimate the density of the expanding plasma and the temperature of the plasma under non-equilibrium ionization (NEI) conditions.  The models employed ionization equilibrium as an assumed initial condition but still depended on NEI calculations when modelling the evolution of the plasma properties out to heights of 2.6~\rsun\ and 3.1~\rsun.

As the plasma's thermal energy evolved, the observationally constrained temperature profile varied in accordance with a given heating parameterization.  Among the five parameterizations that \cite{Wilson.2022} used to define the plasma's rate of heating, one proved to be the most consistent throughout the CME's evolution.  This was found after the parameterizations were applied to the UVCS CME data as well as the data from a similar CME that was constructed by the 3D magnetohydrodynamic (MHD) simulation of the MHD Algorithm outside a Sphere (MAS) code \citep{1999PhPl....6.2217M, 2013Sci...340.1196D, 2013ApJ...777...76L,  2018ApJ...856...75T, Reeves.2019}.  The most consistent heating parameterization produces plasma heating rates that depend on the conservation of magnetic helicity as the CME flux rope expands (in two dimensions akin to a cylinder) and dissipates free magnetic energy \citep{Taylor.1974, Berger.1984, Kumar.1996}.  We use the constrained plasma parameters that derived from this magnetic heating parameterization in order to estimate the flux as if the CME belonged to a distant star.

\subsection{Solar to Stellar Extrapolation}\label{sect: event Wilson extrapolation}

The model temperature profiles of interest were given by \cite{Wilson.2022} in Figure~8 of their work where they assume an inverse square law to express the expansion rate of the plasma.  The evolving density and temperature affects the collisional and radiative excitation rates that yield the model luminosity for the observed plasma.  The model plasma commences its journey at 1.1~\rsun\ and takes 1.3~hours to reach the final UVCS slit height of 3.1~\rsun.  During this journey, we model the luminosity as 
\begin{equation}\label{eq: event Wilson luminosity}
    L_{\odot, CME}\ [\textrm{photons\ s}^{-1}] =  \epsilon_{\lambda} \frac{ M_{\odot, \textrm{CME}} }{ \rho_{\textrm{CME}} }, 
\end{equation}
which incorporates the total emissivity $\epsilon_{\lambda}$ and models its collisonal and radiative excitation rate components as described in the following:
\begin{equation}\label{eq: emissivity}
    \begin{split}
        \epsilon_{\lambda} &= \epsilon_{c,\lambda} + \epsilon_{r, \lambda} \\
        \epsilon_{c,\lambda} = &\ n_{Z,z} \cdot n_e q_{ex, \lambda}(Z,z,T) , \\
        \epsilon_{r,\lambda} = &\ n_{Z,z} \cdot I_\odot(\lambda_i + \delta\lambda_i) \sigma_\lambda \mathcal{W}(r) . \\
    \end{split}
\end{equation}
The total mass density, $\rho_{\textrm{CME}} = m_e n_e + \sum\limits^Z \sum\limits^z m_{Z,z} n_{Z,z}$, is derived from the electron mass density $m_e n_e$, the atomic mass $m_{Z,z}$, and the ion density $n_{Z,z}= n_H A_Z f_{Z,z}$.  The total mass of the CME $M_{\odot, \textrm{CME}}$ is estimated from LASCO white light images.  

For Equation~\ref{eq: emissivity}, the emissivity's collisional component $\epsilon_{c,\lambda}$ depends on the excitation rate coefficient $q_{ex, \lambda}$ for a given spectral line.  The radiative component $\epsilon_{r,\lambda}$ depends the solar disk emission line profile $I_\odot$, which scatters off of the escaping CME plasma at a Doppler redshift $\delta\lambda_i$ from its incident wavelength that corresponds to the speed of the CME.  The scattered radiation also correlates with the absorption cross section $\sigma_\lambda$ and the solid angle $\mathcal{W}(r)$ of the scattering plasma that is subtended by the solar disk.  These plasma parameters were constrained by UVCS observations and subsequently resulted in the evolutionary profiles presented by \cite{Wilson.2022} for the densities, temperatures, and ionization states.  

We evaluate the extrapolated stellar flux of the modelled solar radiation from \ionn{O}{VI} at 1032~\Ang\ and \ionn{C}{III} at 977~\Ang.  The \ionn{O}{VI} and \ionn{C}{III} ions have maximum formation temperatures of 10$^{5.5}$ and 10$^{4.8}$~K, respectively.  A reliable temperature-dependent emission measure distribution is not feasible to construct with the given observations of the plasma at 2.6 and 3.1~\rsun.  Thus, we do not apply an emission measure proportionality to estimate the flux from \ionn{C}{IV}.  An EM distribution would be subject to large uncertainties considering the detected plasma was at heights near 3.0~\rsun, which implies that it is likely subject to NEI conditions, relatively significant contributions from resonant scattering radiation, and subject to frozen-in ionization states.  For \ionn{C}{IV}, we note that its maximum formation temperature of $10^{5.0}$~K is between that of \ionn{O}{VI} and \ionn{C}{III}; consequently, the results from \ionn{O}{VI} and \ionn{C}{III} can act as a qualitative proxy for what the unobserved \ionn{C}{IV} emission might yield.

For an integration time lasting the duration of the modelled 1.3-hour journey, we approximate the stellar CME fluence via Equation~\ref{eq: flux} with $f_\star = 10^2$.  For the \ionn{O}{VI} 1032~\Ang\ line, the constrained model profiles yield the fluences $F_{10\rm{pc},CME} = 50$~photons and $F_{Proxima,CME} = \scinot{3.0}{3}$~photons which suggest that S/N~$\approx$ 5 and 55, respectively.  For the \ionn{C}{III} 977~\Ang\ line, we assume $A_{eff} = 20$~cm$^2$ and find that $F_{10\rm{pc},CME} = 44$~photons and $F_{Proxima,CME} = \scinot{2.6}{3}$~photons which imply that the S/N~$\approx$ 5 and 50, respectively.  These results derive from the observationally constrained physical properties that mostly indicate initially high temperatures of $\gtrsim$10$^6$~K for the CME.  The corresponding temperature profiles decrease to $\lesssim$10$^5$~K when the CME reaches the heights near 3.0~\rsun\ (cf. Figure~8 of \cite{Wilson.2022}).  Since the corresponding densities are greatest at the beginning of the CME's journey, the modelled plasma emissivities are brightest near the coronal surface.  With the corresponding high temperatures there, the observationally constrained emissivities predominantly come from ions and emission lines that are more abundant in very hot environments, unlike the \ionn{O}{VI} and \ionn{C}{III} lines that we have used for these stellar CME approximations.

\begin{figure*}
    \centering
    \captionsetup[subfigure]{position=top, labelfont=bf,textfont=normalfont,singlelinecheck=off,justification=raggedright} 
    \label{fig: signals}
    \subfloat[\label{fig: signals Landi}]{\includegraphics[width=0.33\linewidth, trim=0cm 0cm 0.17in 0.3in, clip]{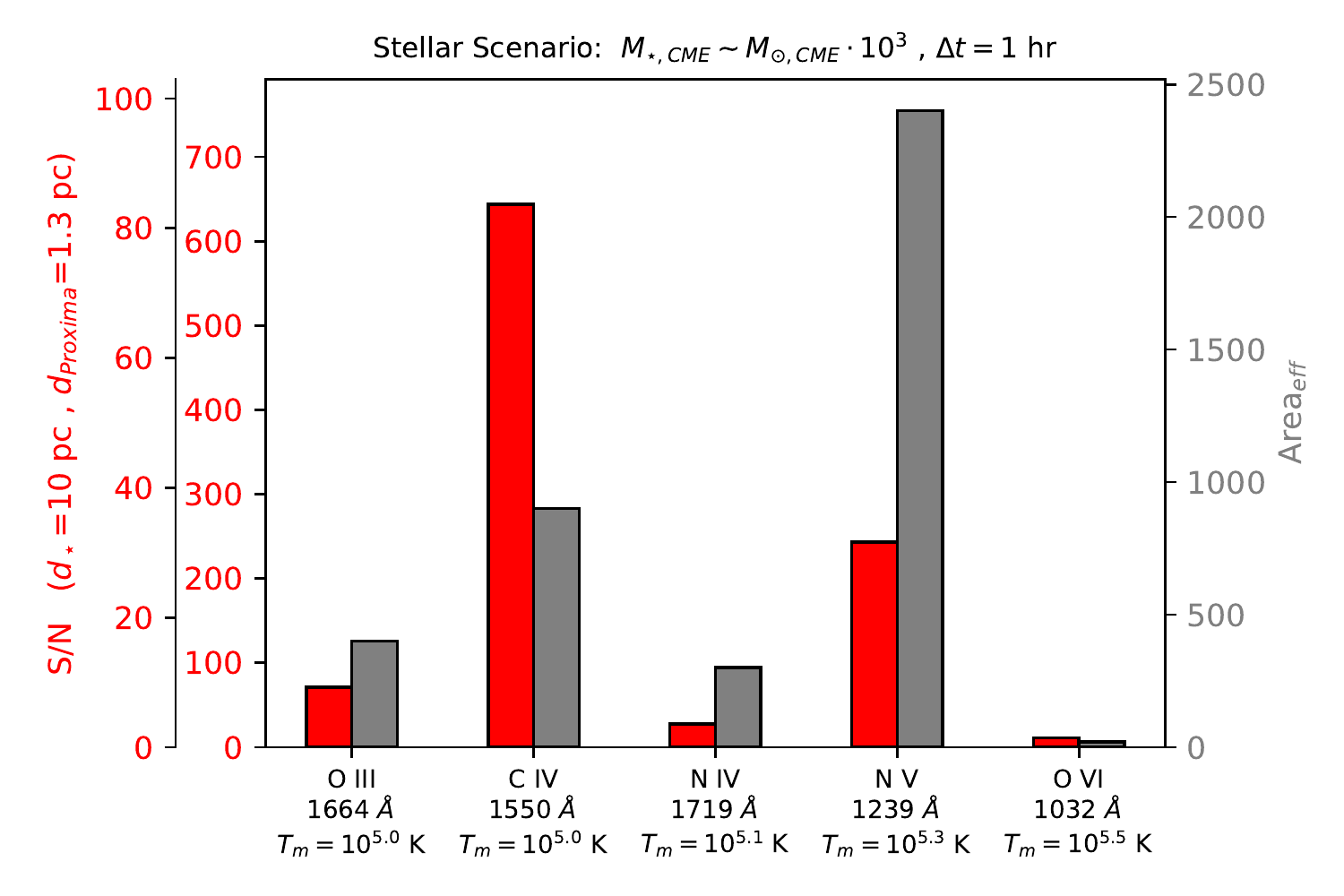}} \hfill
    \subfloat[\label{fig: signals Ciaravella}]{\includegraphics[width=0.33\linewidth, trim=0cm 0cm 0.17in 0.3in, clip]{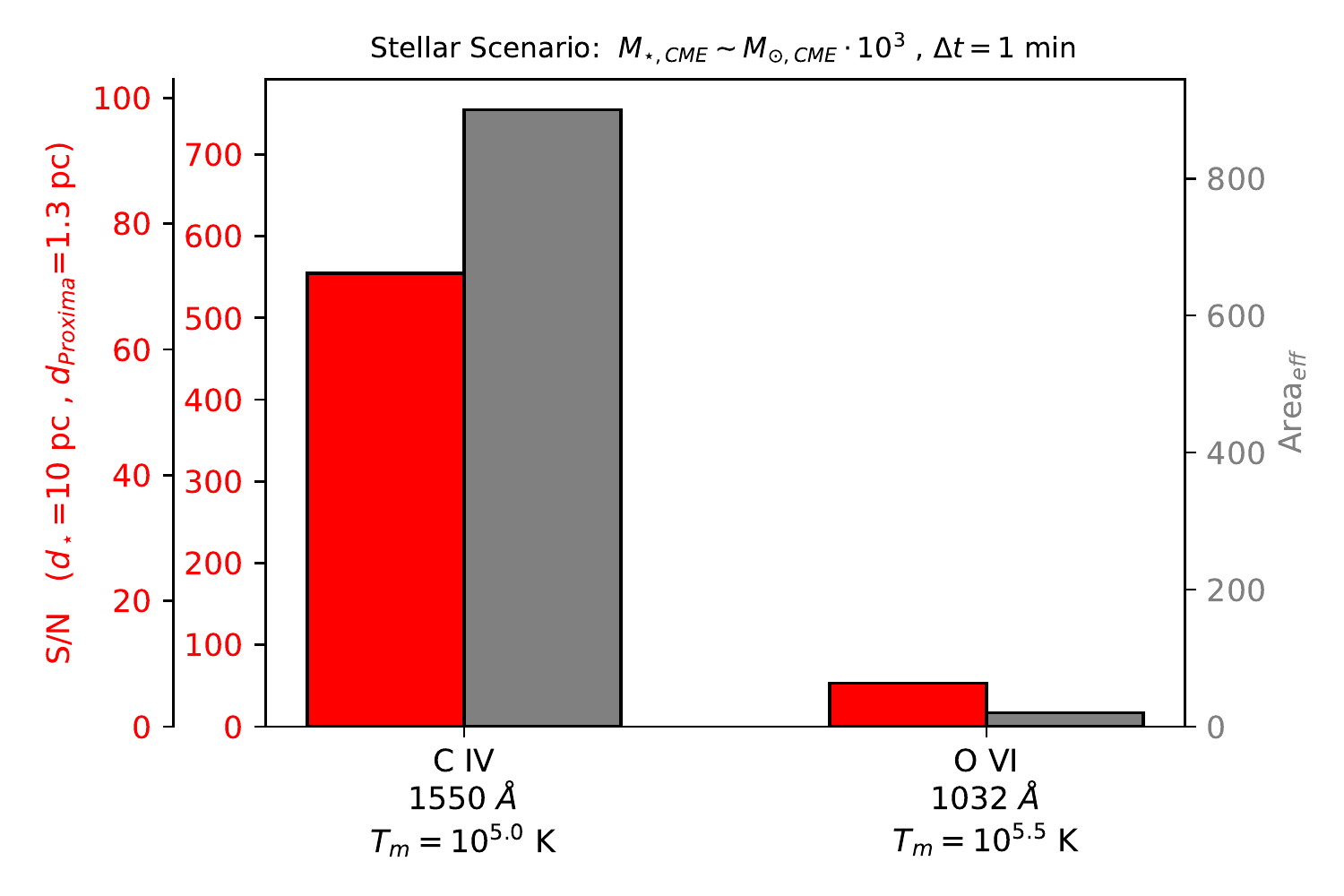}}\hfill
    \subfloat[\label{fig: signals Wilson}]{\includegraphics[width=0.33\linewidth, trim=0cm 0cm 0.17in 0.3in, clip]{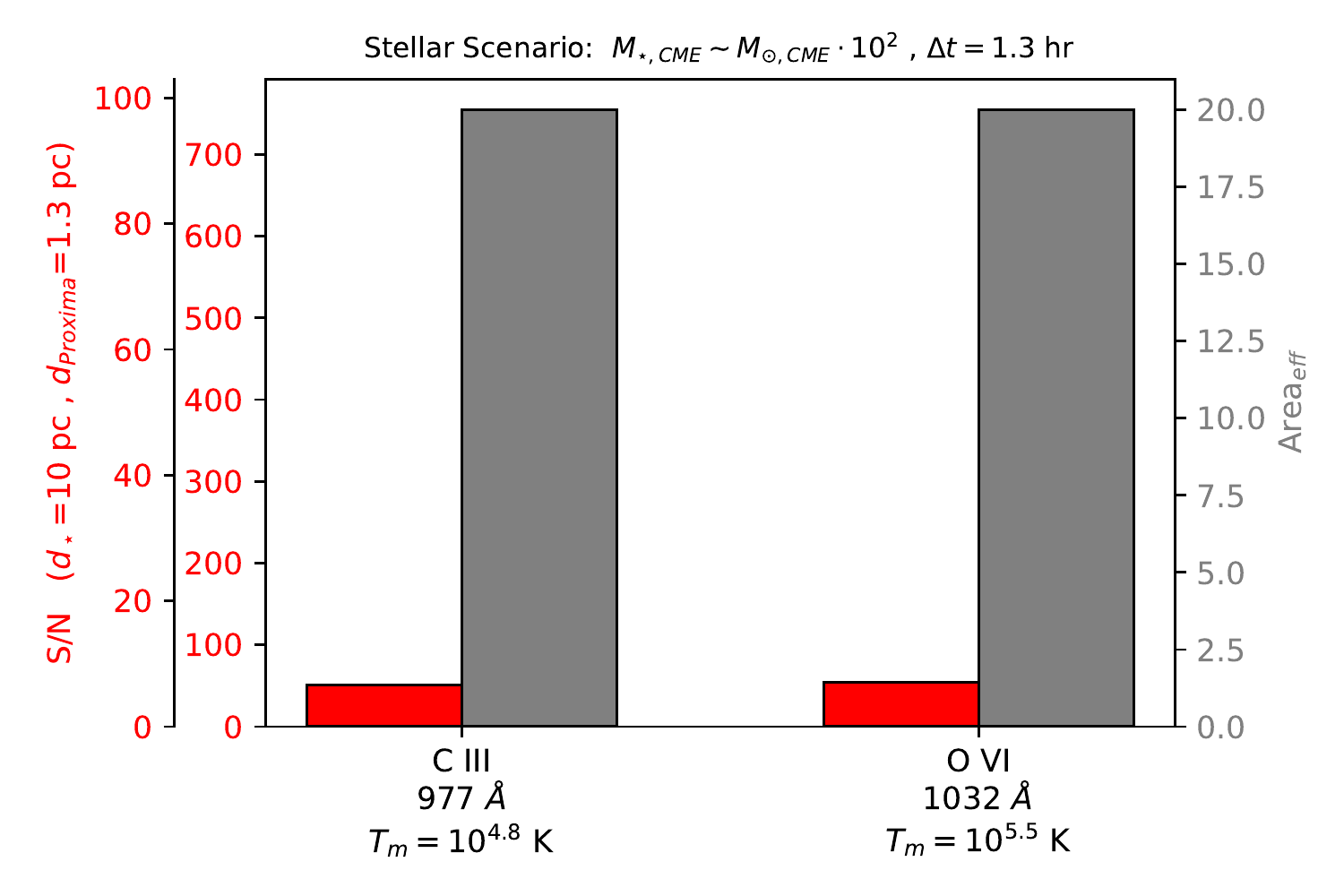}}
    \caption{Predicted emission line signals for potential stellar CMEs, as extrapolated from the diagnostics of solar CMEs studied by \textbf{(a)} \cite{Landi.2010} with an integration time of $\Delta t=1$~hr, \textbf{(b)} \cite{Ciaravella.2000} with $\Delta t=1$~min, and \textbf{(c)} \cite{Wilson.2022} with $\Delta t=1.3$~hr.  The estimated flux for a CME is scaled generically to a distance of 10~pc and specifically to the distance of Proxima Centauri at 1.30~pc. }
\end{figure*}

The \ionn{C}{IV} luminosity is likely to be comparable to those of \ionn{C}{III} and \ionn{O}{VI}, but the effective area is likely to be much larger.  For \textit{HST}/COS, the S/N would be larger by a factor of $\sim$5.  The resultant signal of the blended 1550~\Ang\ line can convey additional diagnostics for the plasma if an instrument's spectral resolution sufficiently distinguishes its doublet lines, at 1548 and 1551~\Ang, from each other.  Due to the collisional and radiative excitation components of the emission lines, the 1551~\Ang\ line can experience radiative pumping.  For a hypothetical stellar CME travelling directly away from (i.e., normal to the surface of) its host star, when its \ionn{C}{IV} ions accelerate to a speed of 580~\unitv\ the chromospheric emission of the \ionn{C}{IV} 1548~\Ang\ line is redshifted to 1551~\Ang\ with respect to the CME's frame of reference.  Consequently, the CME's \ionn{C}{IV} ions will initiate resonant scattering at 1551~\Ang\ which amplifies the total emissivity detected at that wavelength via the line's radiative excitation component. 

Similar to the diagnostic capabilities of the \ionn{O}{VI} doublet lines \citep[\egcite][]{Bemporad.2006, Gilly.2020, Wilson.2022}, the \ionn{C}{IV} doublet can indicate when the stellar CME's speed is near 580~\unitv\ via the measured intensity ratio.  As the CME gains distance away from the host star, each line's radiative component attenuates and the collisional component becomes dominant.  The ratio between the detected lines' intensities becomes similar to the ratio between their collisional emissivities.  The collisional components of the doublet transitions have a ratio~$\approx$~2.0; and, when the radiative pumping of the 1551~\Ang\ line occurs, the total intensity ratio detected deviates from 2.0.  Throughout the duration of an (unresolved) stellar CME observation, the intensity ratio measurements can convey how long the stellar CME maintains a speed of $\sim$580~\unitv\ while the LOS velocity (and acceleration) is determined from the Doppler shift measurements of the spectroscopically resolved doublet.  As a result, the POS component of the velocity and the direction of the stellar CME's propagation can be deduced.

Another property of the \ionn{C}{IV} doublet could have been applied to the stellar UV dimming event of \cite{Ambruster.1986} if the transition lines were resolved.  They interpreted the dimming as the consequence of a stellar CME candidate with relatively cool plasma travelling in front of a hot, bright flare.  Due to the factor of two difference in the \ionn{C}{IV} doublet transitions' collision strengths, the incident chromospheric intensity of the (brighter) 1548~\Ang\ emission line could have been dimmed twice as much as the 1551~\Ang\ emission line.  This likely would have been discerned in the lines' light curves.  Such a detection would have served as additional evidence in the attempt to confirm that the stellar dimming event was due to cool plasma travelling above the corona while absorbing (or scattering) the bright continuum radiation from the hot flare below.    

\section{CME Search Caveats and Conclusion}\label{sect: conclusion}

The search for stellar CMEs is still heavily dependent on the simultaneous occurrence of stellar flares.  Furthermore, \textit{stellar} CME candidates are often confidently proclaimed when their observational signatures are similar to what is predicted from \textit{solar} CME models or what is seen through \textit{solar} CME observations.  The candidates' dependence on stellar flares and solar analogs would not be a contentious point of concern if the plasma diagnostics for each candidate were robustly determined from a variety of observational techniques.  Such techniques can include the concurrent use of several spectroscopic line ratios, where each ratio correlates with a parameter that describes the plasma conditions (e.g., density) and morphology (e.g., expansion speed).

In this work, we have reviewed a few solar CME studies that benefitted from the diagnostic capabilities of ultraviolet emission lines and line intensity ratios.  We propose that such plasma diagnostics can strengthen the credibility of stellar CME detections, as long as the emission lines have a distinguishable signal.  We focused mostly on three emission lines: \ionn{C}{IV} 1550~\Ang, \ionn{O}{VI} 1032~\Ang, and \ionn{C}{III} 977~\Ang.  As summarized in Figure~\ref{fig: signals}, we estimated the brightness of the lines' stellar-based signals, assuming the hypothetical stellar CME radiates similarly to solar CMEs.  Presumably, the radiative profile (and underlying physics) of stellar CMEs would be similar to solar CMEs since the plethora of detected stellar flares often exhibit plasma properties that scale directly with solar flare properties, regardless of the wavelength regime or the unsolar-like radiative energy of stellar superflares ($>10^{33}$~ergs) \citep{Guedel.1996, Aschwanden.2008, Pandey.2012}.

If the physical coupling and statistical association of solar flare-CME events holds true for the stellar events, an efficient search for stellar CMEs might aim at targets that give off high-energy stellar flare signals.  If magnetically active M~dwarfs are the primary targets, their frequent flaring, as well as their abundance in the solar neighborhood, should improve the probability of finding superflares with their associated CMEs.  This high energy stellar flare-CME relationship is qualitatively corroborated by the aforementioned CME candidates from \cite{Houdebine.1990}, \cite{Argiroffi.2019}, and \cite{Namekata.2021}, which were each associated with superflares.  

Despite being a convenient source of observable magnetic activity, caution must be taken when depending on an active flaring star to successfully expel detectable CMEs.  The strength and occurrence rate of such stellar magnetic activity phenomena are attributed partly to the dynamo-generated magnetic fields \citep{Parker.1955, Duvall.1984, Noyes.1984, Wright.2011}.  Ambient fields of high magnetic pressure can confine the plasma that erupts below it.  As a result, the plasma that would have escaped to become a CME reverses its direction and travels back to the coronal floor.  Such failed CMEs might occur frequently on active flaring stars, as long as the kinetic energy of the erupted material is suppressed by the overbearing magnetic field energy \citep{Joshi.2013, Drake.2016, Zuccarello.2017, AlvaradoGomez.2018}.  

Since CMEs need not occur alongside (super)flares, a robust methodology for confirming CME candidates would consistently be capable of constraining the candidate's plasma motions, physical conditions, and energy budget without the simultaneous presence of a flare.  Although counterintuitive, the presence of a flare can introduce scenarios for a false-positive CME detection.  Flare-induced chromospheric brightenings \citep[\egcite][]{Kirk.2017} and chromospheric evaporation \citep[\egcite][]{Milligan.2009} can generate upward plasma flows that yield a Doppler blueshift of several~hundred~\unitv\ or less.  To distinguish the (bright) confined flare plasma from the (faint) escaping CME plasma, a spectrograph that records a stellar flare-CME candidate should have a spectral resolution of 100~\unitv\ or better.  Considering these concerns, the spectroscopic detection and characterization of flare-less stellar CME candidates may be more reliable than detections of stellar flare-CME candidates. 

Furthermore, a flare-less stellar CME candidate may exhibit Doppler redshifts that can be used to confirm its CME identity.  In this case, a detectable redshift would be less likely to derive from the flare-induced plasma downflows of chromospheric condensation \citep{Milligan.2006}.  Instead, this redshift could be the result of the geometry between the CME's propagation direction and limb of the spatially unresolved stellar disk.  As illustrated by \cite{Moschou.2019}, the redshift can derive from a CME erupting near the limb but behind the face of the stellar disk.  As the CME expands and travels at an angle directed partially away from the observer, the plasma can eventually be seen off the limb of the stellar disk.  Since the eruption began behind the face of the star, the possible presence of a concurrent flare would be unknown or impossible to detect directly.  

A survey searching for flare-less stellar CMEs will likely be more successful when observing quiescent Sun-like stars, as opposed to observing magnetically active, young, late-type, low-mass stars that flare frequently.  Despite their energetically weaker and less frequent magnetic activity, stars like our Sun might make for efficient survey targets since their weaker ambient magnetic fields make them less susceptible to failed CME events.  Fortunately, although our Sun successfully launches CMEs almost every day, the planetary magnetosphere and atmosphere of the Earth has not been damaged beyond repair by these CMEs.  By analogy, this bodes well for exoplanets within the habitable zones of CME-launching Sun-like stars.  Furthermore, exoplanets orbiting active M~dwarfs may witness more failed CMEs that are confined to the stellar corona than witness successful CMEs that strip them of their (exo)planetary atmosphere.  These hypotheses can be tested if the search for stellar CMEs coincides with exoplanet surveys.

There is one space-based exoplanet mission concept that might coincidentally detect the type of stellar CME signatures that we have extensively discussed.  The UV-SCOPE (UltraViolet Spectroscopic Characterization Of Planets and their Environments) concept is designed to have a space-based spectrograph sit and stare at transiting exoplanets and their host stars in order to study the upper atmospheres of exoplanets with transmission spectroscopy \citep{Shkolnik.2021, Line.2021, Loyd.2021}.  Its primary mirror will have a diameter of 60~cm.  Its spectral coverage will span the range 1205--4000~\Ang\ with a corresponding range in resolving power of $R\sim$~6000--100 from the short to long wavelength range\footnote{The spectrum includes the doublets of \ionn{C}{IV} and \ionn{Mg}{II} near 1550 and 2800~\Ang, respectively, which were studied by \cite{Ambruster.1986} for stellar UV dimming event.}.  As a dedicated observatory, UV-SCOPE will apply long temporal coverage (per target) so that the impact of stellar UV emission on exoplanets can be examined in detail.  As a bonus, its temporal and spectral coverage may suffice in detecting and characterizing stellar CME candidates with diagnostic UV emission lines.

\vspace{10pt}

We acknowledge that this work has benefitted from the use of NASA's Astrophysics Data System.  SOHO is a mission from the joint collaboration of ESA and NASA.  This work was supported by NASA Grant 80NSSC19K0853 to the Smithsonian Astrophysical Observatory.



\end{document}